\newcommand{\CLASSINPUTtoptextmargin}{19.1mm}
\newcommand{\CLASSINPUTbottomtextmargin}{19.1mm}
\newcommand{\CLASSINPUToutersidemargin}{19.1mm}

\documentclass[conference,letterpaper]{IEEEtranm}

\IEEEoverridecommandlockouts

\usepackage{subfiles}
\usepackage{excludeonly}

\usepackage[utf8]{inputenc}
\usepackage[T1]{fontenc}
\usepackage[english]{babel}

\usepackage{lmodern}

\usepackage{mathtools}
\usepackage{amssymb} 

\usepackage{bm} 

\usepackage{pifont}

\usepackage[version=3]{mhchem} 
\usepackage{esint} 
\usepackage{dsfont} 
\usepackage[makeroom]{cancel} 
\usepackage{empheq}
\usepackage{physics}
\usepackage[binary-units=true]{siunitx}

\usepackage{graphicx}
\usepackage{float}
\makeatletter
\let\MYcaption\@makecaption
\makeatother

\usepackage[font=footnotesize]{subcaption}

\makeatletter
\let\@makecaption\MYcaption
\makeatother

\usepackage{booktabs}
\usepackage{makecell}
\usepackage{threeparttable}
\usepackage{multirow}
 
\usepackage{enumitem}

\usepackage{algorithm}
\usepackage{algpseudocode}
\usepackage{listings}

\usepackage[dvipsnames]{xcolor}

\usepackage{tcolorbox} 

\usepackage{varioref} 
\PassOptionsToPackage{hyphens}{url}
\usepackage[hidelinks,breaklinks=true,linktoc=all]{hyperref} 
\hypersetup{colorlinks=true,
allcolors=black} 
\usepackage[nameinlink,capitalise,noabbrev]{cleveref} 
\crefformat{equation}{(#2#1#3)}
\Crefformat{equation}{Equation (#2#1#3)}
\crefrangeformat{equation}{(#3#1#4) to~(#5#2#6)}
\crefmultiformat{equation}{(#2#1#3)}%
{ and~(#2#1#3)}{, (#2#1#3)}{ and~(#2#1#3)}
\Crefmultiformat{equation}{Equations (#2#1#3)}%
{ and~(#2#1#3)}{, (#2#1#3)}{ and~(#2#1#3)}

\crefname{appendix}{appendix}{appendices}
\Crefname{appendix}{Appendix}{Appendices}

\usepackage[noadjust]{cite} 

\usepackage{pgf,tikz,pgfplots}
\pgfplotsset{compat=1.10}
\usetikzlibrary{shapes,arrows,calc,arrows.meta,
patterns,backgrounds,positioning,fit}
\tikzset{every picture/.style=black}
\usepackage{circuitikz}

\tikzset{three sided/.style={
        draw=none,
        append after command={
            [shorten <= -0.5\pgflinewidth]
            ([shift={(-1.5\pgflinewidth,-0.5\pgflinewidth)}]				  \tikzlastnode.north east)
        edge([shift={( 0.5\pgflinewidth,-0.5\pgflinewidth)}]				  \tikzlastnode.north west) 
            ([shift={( 0.5\pgflinewidth,-0.5\pgflinewidth)}]				  \tikzlastnode.north west)
        edge([shift={( 0.5\pgflinewidth,+0.5\pgflinewidth)}]				  \tikzlastnode.south west)            
            ([shift={( 0.5\pgflinewidth,+0.5\pgflinewidth)}]				  \tikzlastnode.south west)
        edge([shift={(-1.0\pgflinewidth,+0.5\pgflinewidth)}]				  \tikzlastnode.south east)
        }
    }
}

\usepackage{psfrag}
\usepackage{pstool} 

\usepackage{lipsum}  
\usepackage{cuted}
\usepackage{stfloats}

\makeatletter
\newcommand{\getfontsize}{\f@size pt}
\makeatother

\newcommand*\ds{\displaystyle}

\newcommand*\diff{\mathop{}\!\mathrm{d}}

\newcommand*\kB{k}

\newcommand*\Vth{V_{\mathrm{th}}}

\newcommand*\VG{V_{\mathrm{G}}}

\newcommand*\vGS{v_{\mathrm{GS}}}
\newcommand*\vDS{v_{\mathrm{DS}}}

\newcommand*\VDD{V_{\mathrm{DD}}}

\newcommand{\dt}{\diff{t}}

\newcommand*\vIN{v_{\mathrm{IN}}}

\renewcommand*\tr{t_{\mathrm{r}}}

\newcommand*\vC{v_{\mathrm{OUT}}}
\newcommand*\Rav{\overline{R}}
\newcommand*\Gav{\overline{G}}
\newcommand*\Dt{\Delta t}
\newcommand*\Dq{\Delta q}
\newcommand*\Dv{\delta v}
\newcommand*\EIN{E_{\mathrm{IN}}}
\newcommand*\Edissip{E_{\mathrm{dissip}}}
\newcommand*\EC{E_{\mathrm{bit}}}
\newcommand*\Edissipmin{E_{\mathrm{dissip,min}}}

\definecolor{k}{rgb}{0 0 0}
\definecolor{r}{rgb}{1 0 0}
\definecolor{g}{rgb}{0 1 0}
\definecolor{b}{rgb}{0 0 1}
\definecolor{orange}{rgb}{1,0.7,0}
\definecolor{c}{rgb}{0 1 1}
\definecolor{cc}{RGB}{64 224 208}
\definecolor{m}{rgb}{1 0 1}
\definecolor{khaki}{RGB}{128 128 0}
\definecolor{deepskyblue}{RGB}{0 191 255}
\definecolor{darkMagenta}{rgb}{0.5 0 0.5}
\definecolor{chocolateBrown}{RGB}{98 52 18}
\definecolor{lightBrown}{RGB}{189 154 122}
\definecolor{mybrown}{RGB}{127 37 0}
\definecolor{bordeaux}{RGB}{131 41 85}
\definecolor{myGreen}{RGB}{134,180,44}
\definecolor{gray_gate}{RGB}{211,208,205}
\definecolor{yellow_oxide}{RGB}{244,231,164}

\definecolor{h}{rgb}{0 0 0}

\definecolor{l}{rgb}{0 0 1}

\graphicspath{{figures/}}

\begin{document}

\title{
\raisebox{6.3mm}{\strut}
The non-Landauer Bound for the Dissipation of Bit Writing Operation 
\thanks{The work has been supported by the Research Project "Thermodynamics of Circuits for Computation" of the National Fund for Scientific Research (F.R.S.-FNRS) of Belgium.}
}

\author{
\IEEEauthorblockN{Léopold Van Brandt and Jean-Charles Delvenne}
\IEEEauthorblockA{
		ICTEAM, UCLouvain, Belgium\\
       {\tt leopold.vanbrandt@uclouvain.be}
      			 }
       }
       
\maketitle

\begin{abstract}

We propose a novel bound on the minimum dissipation required in any circumstances to transfer a certain amount of charge through any 
resistive device. We illustrate it on the task of writing a logical 1 (encoded as a prescribed voltage) into a capacitance, through various linear or nonlinear devices. We show that, even though the celebrated Landauer bound (which only applies to bit erasure) does not apply here, one can still formulate a “non-Landauer” lower bound on dissipation, that crucially depends on the time budget to perform the operation, as well as the average conductance of the driving device. We compare our bound with empirical results reported in the literature and realistic simulations of CMOS pass and transmission gates in decananometer technology. 
Our non-Landauer bound turns out to be a
quantitative benchmark to assess the (non-)optimality of a writing operation.  

\end{abstract}


\section{Introduction}

Landauer's bound \cite{Landauer1961,Izydorczyk2009,Parrondo2015} states that an erasure of a bit, stored in an electronic device or any other type of physical memory, must dissipate at least $\kB T \ln 2$ (where $\kB$ is Boltzmann's constant and $T$ is the temperature). This figure  cannot be taken as a benchmark for the dissipation of any kind of operation in digital circuits, however. Indeed it does not apply to logically reversible operations such as writing (or switching or copying) a bit into an empty memory, e.g. bringing the state of the memory from a logical 0 to a logical 1. Sub-$\kB T$ bit-switching operations have been for instance discussed and carried out experimentally by \cite{Orlov2012,Snider2012}, exhibiting a trade-off between speed and dissipation of the operation. 

While an arbitrarily low dissipation can indeed be theoretically achieved by implementing an arbitrarily slow process, here we show that, for any linear or nonlinear driving 
	device
and any specified time $\Dt$ for the bit writing operation, there \emph{is} a fundamental lower bound, yet “non-Landauer”, which predicts the minimum dissipation than can be achieved with the optimal protocol. Our bound is not limited to bit writing process but can assess any operation 
involving
a charge flow through 
dissipative
devices.

 In \cref{section:Problem Statement and Main Result}, we define the bit writing operation, first generally and then particularized for a capacitance load, and we present a novel lower bound for the energy dissipation of the process. In \cref{section:Applications} we compare our bound with experimental results \cite{Orlov2012} and simulations of CMOS devices. We show for example that for the chosen parameters, an nMOS pass gate is two decades away from optimality, while a transmission gate is only a factor two over the bound, thereby illustrating the 
scope 
of our bound as a benchmark of efficiency. Conclusions are in \cref{Discussion and Conclusions}.

\section{Problem Statement and Main Result}
\label{section:Problem Statement and Main Result}

\newcommand\myfontsize{\normalsize}
\newcommand\dx{0.75}
\newcommand\xspacing{1.5}
\begin{figure}
\centering
\begin{circuitikz}[european, american voltages, transform shape, line cap=rect, nodes={line cap=butt},scale=1]
\draw[draw=none,opacity=0]
(0,0) node[nmos,rotate = -90,yscale=-1] (nMOS) {};
\draw
(nMOS.D) to[R=Nonlinear device,-] (nMOS.S);
\draw 
(nMOS.D) -- ++(-\xspacing*\dx,0) node[] (IN) {};
\draw 
(IN) ++ (0,-2*\dx) node[ground,scale=1,color=black] (GNDIN) {}
to[V=\textcolor{b}{$\vIN(t)$},invert] (IN);
\draw 
(nMOS.S) ++(\xspacing*\dx,0) node[label={[font=\myfontsize]right:$\textcolor{r}{\vC(t)}$}] (OUT) {}
(nMOS.S) to[short,-*] (OUT);
\draw
(OUT)++(0,-2*\dx) node[ground,scale=1,color=black] {} to[C,color=black,l_={$C$}] (OUT);
{
\color{mybrown}
\ctikzset{current/distance = 0.5}
\draw 
(nMOS.D)++(0,-0.5*\dx) node[] (nDi) {}
(nMOS.S)++(0,-0.5*\dx) node[] (nSi) {};
\draw[->] 
(nDi) -- node[below, color=mybrown, font=\myfontsize] {$i(t)$} (nSi);
}
\end{circuitikz}
\caption{Charging a (possibly non-constant) capacitance through some (possibly nonlinear) dissipative dipole.}
\label{fig_statement}
\end{figure}
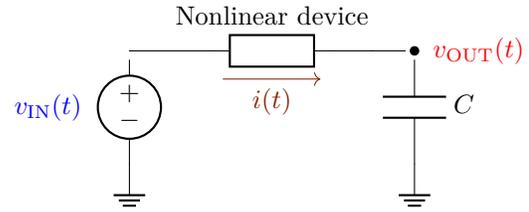

With a view to a large class of nonlinear physical systems made of both dissipative and energy-storing components, we are most generally interested in \emph{charge transfer} from one circuit (referred to as the “input”) to another (“output”) through a driving device.
In the schematic depicted in \cref{fig_statement}, the charges are supplied by an ideal voltage source while the load is a constant capacitor, for illustrative purposes and without loss of generality at this stage. 
The transmitting dipole has a finite conductance, thus is \emph{dissipative}, but is otherwise arbitrary: 
it may be a linear resistor (modelling for instance the path losses) or a CMOS gate.

The storage circuit exploits the charge $q(t)$ as state variable to represent a logical 0 ($q(t) = q_0 = 0$, by convention)
or a logical 1 ($q(t) = q_1$, positive).
We focus on the bit writing operation, i.e. switching from 0 to 1 through a charge transfer.
Our goal is thus to bring a charge $\Dq$ on the capacitor in a specified time $\Dt$, however by minimizing the \emph{energy dissipation} as much as possible.
The $\Dt$ will sometimes be referred to as \emph{write time} or \emph{response time} and obviously depends on the input signal $\textcolor{k}{\vIN(t)}$, the behaviour of the transmission device and on the capacitive load.
For a given circuit, 
an imposed
$\Dt$ therefore partly determines the properties of the input signal, notably its speed.
There might be additional technological constraints 
restricting
the shape of the input signal (maximum supply voltage, finite slew rate,...).

\subsection{Definitions of the Important Physical Quantities}


In \cref{fig_statement}, $\vIN(t)$ and $\vC(t)$ denote the input and output node voltages, respectively, and
\begin{equation}
\label{eq:Dv}
\Dv(t) \equiv \textcolor{k}{\vIN(t)} - \textcolor{k}{\vC(t)}
\text{,}
\end{equation}
the voltage difference across the dipole.
Let $i(t)$ be the electrical current flowing through the transmission device.
The charge transferred from the input to the output during a time interval $\Dt$ is
\begin{equation}
\label{eq:Dq}
\Dq \equiv \int_{t_0}^{t_0 \mathrlap{+ \Dt}} \diff{q}
= \int_{t_0}^{t_0 + \Dt} \! \dv{q(t)}{t} \, \dt
= \int_{t_0}^{t_0 \mathrlap{+ \Dt}} i(t) \, \dt
\text{.}
\end{equation}
For the special case of a constant capacitance output load (\cref{fig_statement}), we have 
$i(t) = C\diff{\vC(t)}/\diff{t}$
and

\begin{equation}
\label{eq:Dq C}
\Dq 
= C \, \big( \vC (t_0 + \Dt) - \vC (t_0) \big) \\
= C \, V_1
\text{,}
\end{equation}
$V_1$ being the prescribed voltage defining the logic level 1.


It is instructive to carry out the \emph{energy balance} of the circuit. The energy supplied at the input splits, 
as a result of
\eqref{eq:Dv}, into two components:
\begin{equation}
\EIN = 
\Edissip + \EC
\text{.}
\end{equation}
\begin{equation}
\label{eq:EC}
\EC = \int_{t_0}^{t_0 \mathrlap{+ \Dt}} \vC(t) \, i(t) \, \dt
\end{equation}
is the reversible component of the supplied energy, 
“profitably” transferred to the storing element. For the special case of the constant $C$, we of course retrieve
\begin{equation}
\label{eq:EC C}
\EC = \frac{1}{2} C V_1^2 = \frac{\Dq^2}{2 \,C}
\text{.}
\end{equation}
Thereafter, we will most often refer to this quantity \eqref{eq:EC C} as the \emph{bit energy}, following \cite{Orlov2012}'s terminology.

The \emph{energy dissipation} of the switching operation is 
\begin{equation}
\label{eq:Edissip}
\Edissip 
= \int_{t_0}^{t_0 \mathrlap{+ \Dt}} \Dv(t) \, i(t) \, \dt
\text{.}
\end{equation} 
Only a device with an infinite conductance ($i(t)/\Dv(t) = \infty$) is able to transmit the charges (finite non-zero $i(t)$) while keeping $\Dv(t) = 0$, which would result in $\Edissip = 0$ according to \eqref{eq:Edissip}.
This theoretical limit case actually corresponds to a simple lossless (zero resistance) connecting wire between the source and the storing element.




\subsection{Lower Bound for the Dissipation}

We may consider a yield coefficient assessing the efficiency of the charge/energy transfer in the circuit of \cref{fig_statement}:
\begin{equation}
\label{eq:EC/EIN}
\frac{\EC}{\EIN} = \frac{\EC}{\EC + \Edissip} = \frac{1}{1 + \Edissip/\EC}
\text{,}
\end{equation}
which approaches $1$ as the \emph{loss ratio} $\Edissip/\EC$ (dissipation normalized by the bit energy) decreases.
The main result of this paper is a general lower bound on $\Edissip$:
\begin{equation}
\label{eq:Edissipmin}
\Edissip \geq \Edissipmin 
\equiv \frac{\Dq^2}{\Gav \, \Dt} \\
= \frac{\ds\bigg( \int_{t_0}^{t_0 \mathrlap{+ \Dt}} i(t) \, \dt \bigg)^2}{\Gav \, \Dt}
\text{,}
\end{equation}
where
\begin{equation}
\label{eq:Gav}
\Gav \equiv \frac{1}{\Dt} \int_{t_0}^{t_0 + \Dt} \! \frac{i(t)}{\Dv(t)} \, \dt
\end{equation}
may be interpreted as the \emph{average} conductance of the nonlinear driving device over the whole charging time interval.
The derivation is proposed in \Cref{appendix:Edissipmin}, based on Cauchy-Schwarz inequality. 
The formulation \eqref{eq:Edissipmin} is general in the sense that it does not depend on the output load (whose a constant $C$ is a special important case). 

The bound is tight, i.e. the bit writing operation is energetically optimal, if the voltage difference across the nonlinear device is constant (\Cref{appendix:Edissipmin}):
\begin{equation}
\label{eq:Dv(t) cst}
\Dv(t) = \Delta V = \text{constant}
\text{.}
\end{equation}
Remarkably, the condition \eqref{eq:Dv(t) cst} is general in the sense that it does not rely on any assumption regarding the transmission device of \cref{fig_statement}.
It is important to acknowledge that, despite its apparent simplicity, the condition \eqref{eq:Dv(t) cst} can be hard to translate into a 
optimal $\vIN(t)$ in practical switching circuits exhibiting substantial nonlinearities (see CMOS applications in \cref{subsection:CMOS}).
The shape of the $\vIN(t)$ such that \eqref{eq:Dv(t) cst} is satisfied can indeed be very complex to implement.

For the important special case of a constant output capacitance 
(see \cref{eq:Dq C}), \eqref{eq:Edissipmin}  expands as
\begin{equation}
\label{eq:Edissipmin C}
\Edissip \geq \frac{1}{2} C V_1^2 \, \frac{2 C}{\Dt \, \Gav}
\text{.}
\end{equation}
That is, the loss ratio is bounded by the following relation:
\begin{equation}
\label{eq:Edissip/EC C}
\frac{\Edissip}{\EC} \geq \frac{\Edissipmin}{\EC} = \frac{2 C}{\Dt \, \Gav}
\text{.}
\end{equation}
In \eqref{eq:Edissip/EC C}, the quantity $C/\Gav$ reminds us the time constant of a linear $RC$ circuit (discussed in detail below in \cref{subsection:RC}), however extended to a general nonlinear circuit through the definition \eqref{eq:Gav}.
\Cref{eq:Edissip/EC C} tells us that the dissipated energy decreases as the write time $\Dt$, whose 
value is relative to the time constant of the circuit, increases. This trend is the classical speed/dissipation tradeoff \cite{Athas1994,Snider2012}.
We have provided \cref{eq:Edissipmin C,eq:Gav} to predict the minimum dissipation, as well as the optimal protocol \eqref{eq:Dv(t) cst} to achieve it.

\section{Applications}
\label{section:Applications}

In this section, we exploit our fundamental lower bound to discuss the energy efficiency of several practical switching electronic circuits.

\subsection{Linear $RC$ circuit}
\label{subsection:RC}

\begin{figure}
\centering
\captionsetup[subfigure]{skip=2pt}
\begin{subfigure}[t]{3cm}
\centering
\psfragscanon
\footnotesize
\psfrag{0}[cc][cc]{$0$}
\psfrag{0.5}[cc][cc]{$0.5$}
\psfrag{0.8}[cc][cc]{$0.8$}
\psfrag{t0}[cc][cc]{$t_0$}
\psfrag{t0+tr}[cc][cc]{$t_0 + \tr$}
\psfrag{VDD}[cr][cr]{\small $\VDD$}
\psfrag{tr}[tl][tl]{\small  $\textcolor{k}{\tr}$}
\includegraphics[scale=1]{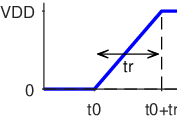}
\caption{}
\label{fig_ramp}
\end{subfigure}%
~
\begin{subfigure}[t]{5.8cm}
\centering
\begin{circuitikz}[american voltages, transform shape, line cap=rect, nodes={line cap=butt},scale=0.9]
\draw[draw=none,opacity=0]
(0,0) node[nmos,rotate = -90,yscale=-1] (nMOS) {}
;
\draw
(nMOS.D) to[R=$R$,-] (nMOS.S);
\draw 
(nMOS.D) -- ++(-\xspacing*\dx,0) node[] (IN) {};
\draw 
(IN) ++ (0,-2*\dx) node[ground,scale=1,color=black] (GNDIN) {}
to[V=\textcolor{b}{$\vIN(t)$},invert] (IN);
\draw 
(nMOS.S) ++(\xspacing*\dx,0) node[label={[font=\myfontsize]above:\textcolor{r}{$\vC(t)$}}] (OUT) {}
(nMOS.S) to[short,-*] (OUT);
\draw
(OUT)++(0,-2*\dx) node[ground,scale=1,color=black] {} to[C,color=black,l_={$C$}] (OUT);
{
\color{mybrown}
\ctikzset{current/distance = 0.5}
\draw 
(nMOS.D)++(0,-0.5*\dx) node[] (nDi) {}
(nMOS.S)++(0,-0.5*\dx) node[] (nSi) {};
\draw[->] 
(nDi) -- node[below, color=mybrown, font=\myfontsize] {$i(t)$} (nSi);
}
\end{circuitikz}
\caption{}
\label{fig_RC}
\end{subfigure}
\caption{\subref{fig_ramp} Linear input ramp. \subref{fig_RC} Linear $RC$ circuit.}
\label{fig_ramp_RC}
\end{figure}

The experimental setup of \cite{Orlov2012}, intended to prove that it is possible to write a bit with a sub-$\kB T$ dissipation, can be modelled as depicted in \cref{fig_ramp_RC}: a constant capacitor ($C = \SI{100}{\pico\farad}$) is charged by a voltage ramp through a series resistor ($R = \SI{1.1}{\kilo\ohm}$).
Whereas \cite{Orlov2012} only describes the input signal as “smooth and gradual”, we reasonably assume a linear ramp (\cref{fig_ramp}) \cite{Snider2012}.
We will show that, for the speed regime studied by \cite{Orlov2012}, the exact shape of the signal has a negligible influence on the experimental results.

This preliminary case study can be covered in detail analytically.
Subsequently, the \emph{exact} dissipation is computed according to \eqref{eq:Edissip}, for some response time $\Dt$.  

In the most general case, we must distinguish $V_1$, the voltage defining the logic level 1, from $\VDD$, the \emph{supply voltage} corresponding here to the final amplitude of the voltage ramp.
For a non-zero $R$, we have necessarily $\vC(t) < \vIN(t)$, meaning that $\vC(t)$ never reaches exactly $\VDD$. 
$V_1$ must therefore be suitably defined, for instance as a fraction of $\VDD$, that is $V_1 = \alpha \VDD$ with $\alpha$ close to 
unity.
Thus, most generally, the output response time $\Dt$ differs from the rise time of the input ramp ($\tr$): both situations $\Dt < \tr$ and $\Dt \gg \tr$ can occur, depending on the choice of $\alpha$ and the speed of the input ramp relative to the $RC$ time constant.
If the $RC$ dynamics is negligible, $V_1 \approx \VDD$ can be rigorously assumed.

\subsubsection{Quasi-Static Case}

We first focus on the conditions 
selected by \cite{Orlov2012} for a sub-$\kB T$ measurement.
We compute $RC = \SI{110}{\nano\second}$, while $\tr$ ranges between $\SI{64}{\micro\second}$ and $\SI{640}{\micro\second}$ \cite{Orlov2012}. 
The quasi-static condition (“slow” input and “slow” bit writing operation) $\tr \gg RC$ is comfortably verified, which implies that $\vC(t)$ closely follows $\vIN(t)$, with almost no delay. We emphasize that $\Dv(t)$ is small, yet non-zero and measurable (down to the $\si{\nano\volt}$ \cite{Orlov2012,Snider2012}).
We may consider $V_1 \approx \VDD$ ($\alpha \approx \SI{100}{\percent}$) together with $\Dt \approx \tr$.

\begin{figure}
\small
\centering
\psfragscanon
\psfrag{tr [us]}[cc][cc]{$\tr\, [\si{\micro\second}]$}
\psfrag{Edissip/kT}[cc][cc]{$\Edissip/\kB T$}
\psfrag{e2}[cr][cr]{$10^{2}$}
\psfrag{e1}[cr][cr]{$10^{1}$}
\psfrag{e0}[cr][cr]{$10^0$}
\psfrag{em1}[cr][cr]{$10^{-1}$}
\psfrag{em2}[cr][cr]{$10^{-2}$}
\psfrag{64}[cc][cc]{$64$}
\psfrag{256}[cc][cc]{$256$}
\psfrag{640}[cc][cc]{$640$}
\psfrag{Orlov (exp)}[cl][cl]{\color{k}Experimental \cite{Orlov2012}}
\psfrag{Lower bound}[cl][cl]{\color{deepskyblue}Fundamental lower bound}
\psfrag{"Landauer kT.ln(2)"}[cl][cl]{\color{r}“Landauer’s  $\kB T \log 2$ bound”}
\includegraphics[scale=1]{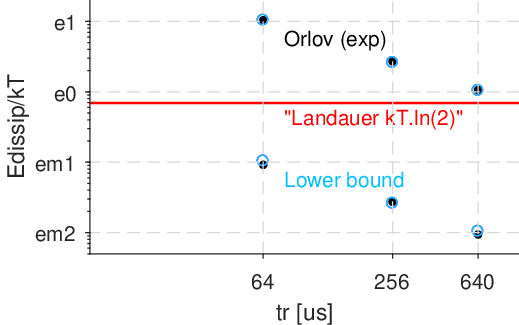}
\caption{Dissipated energy versus switching time for the $RC$ circuit in quasi-static conditions, two different $V_1 = \SI{500}{\micro\volt}$ and $\SI{50}{\micro\volt}$ (bit energies are $\EC = 3000 \kB T$ and $\EC = 30 \kB T$, respectively) \cite{Orlov2012}.}
\label{fig_RC_Orlov}
\end{figure}

In this speed regime, the exact dissipation can be shown to
tightly match the proposed lower bound \cref{eq:Edissipmin C} that takes the simplified form
\begin{equation}
\label{eq:Edissip RC tr << RC}
\Edissip \approx \Edissipmin = \EC \, \frac{2 RC}{\Dt}
\text{,}
\end{equation}
since $\Gav = 1/R = \text{constant}$.
While the lower bound, in a general context \eqref{eq:Edissipmin}\eqref{eq:Edissip/EC C} as well as in the linear case, is a novel contribution to our knowledge, we should mention  that the dissipation in quasi-static conditions (right-hand side of \eqref{eq:Edissip RC tr << RC}) was already obtained earlier \cite{Athas1994,Ye1996,Snider2012,Frank2020}.
Furthermore, we show in \cref{fig_RC_Orlov} that the predictions are in excellent agreement with \cite{Orlov2012}'s experimental results, up to small discrepancies, which we attribute to the non-idealities of the measurement \cite{Orlov2012,Snider2012}. 
The combined experimental, analytical modelling and theoretical prediction efforts comforts us in a major conclusion: 
although there is no such a thing as “Landauer’s  $\kB T \log 2$ bound” to write a bit, there is a fundamental minimum dissipation, precisely given by \eqref{eq:Edissip RC tr << RC} for an $RC$ circuit in slow conditions.
\Cref{eq:Edissipmin} generalizes it for the wide class of nonlinear devices, whose adiabatic logic gates \cite{Athas1994,Ye1996,Snider2012,Frank2020} or conventional CMOS gates are examples.



\subsubsection{Faster Bit-Switching Case and Optimal Protocol}

\begin{figure}
\centering
\psfragscanon
\small
\psfrag{0}[cc][cc]{$0$}
\psfrag{t0}[cc][cc]{$t_0$}
\psfrag{t0+DT}[cc][cc]{$t_0 + \Dt$}
\psfrag{V1 + DV}[cr][cr]{\small $V_1 + \Delta V$}
\psfrag{V1}[cr][cr]{\small $V_1$}
\psfrag{DV}[cr][cr]{\small  $\textcolor{k}{\Delta V}$}
\psfrag{vIN(t)}[tl][tl]{\small $\textcolor{deepskyblue}{\vIN(t)}$}
\psfrag{vC(t)}[tl][tl]{\small $\textcolor{r}{\vC(t)}$}
\includegraphics[scale=1]{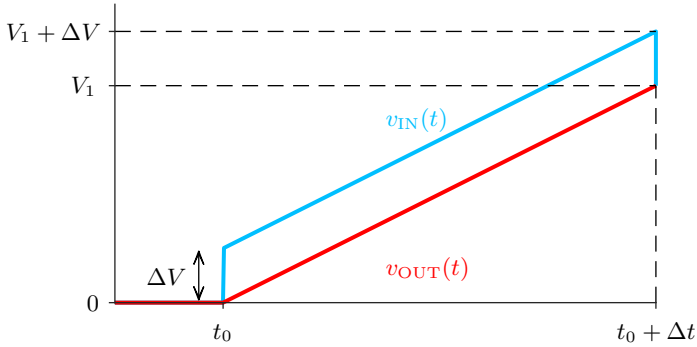}
\caption{Optimal protocol to achieve the minimum dissipation for the linear $RC$ circuit.}
\label{fig_RC_affine}
\end{figure}

We move back to our general problem stated in \cref{section:Problem Statement and Main Result}: given an imposed response time $\Dt$ (for $\vC$ to reach at least $V_1$) can we find an optimal $\vIN(t)$ so as to minimize the energy dissipation?
The answer is yes. It can be shown by direct calculation that the  \emph{affine} $\textcolor{k}{\vIN(t)}$ sketched in \cref{fig_RC_affine} satisfies the optimality condition \eqref{eq:Dv(t) cst}, where $\Delta V$ must be chosen as a function of $V_1$ and $\Dt$:
\begin{equation}
\label{eq:RC affine DV}
\Delta V = V_1 \frac{RC}{\Dt}
\text{.}
\end{equation}


\begin{figure}
\small
\centering
\psfragscanon
\psfrag{DT/RC}[cc][cc]{$\Dt/RC$}
\psfrag{Edissip/(C.V1^2/2)}[cc][cc]{$\Edissip/\EC$}
\psfrag{e2}[cc][cc]{$10^{2}$}
\psfrag{e1}[cc][cc]{$10^{1}$}
\psfrag{e0}[cc][cc]{$10^0$}
\psfrag{0}[cc][cc]{$0$}
\psfrag{0.1}[cc][cc]{$0.1$}
\psfrag{0.2}[cc][cc]{$0.2$}
\psfrag{0.3}[cc][cc]{$0.3$}
\psfrag{0.4}[cc][cc]{$0.4$}
\psfrag{0.5}[cc][cc]{$0.5$}
\psfrag{0.6}[cc][cc]{$0.6$}
\psfrag{0.7}[cc][cc]{$0.7$}
\psfrag{0.8}[cc][cc]{$0.8$}
\psfrag{0.9}[cc][cc]{$0.9$}
\psfrag{1}[cc][cc]{$1$}
\psfrag{1.5}[cc][cc]{$1.5$}
\psfrag{Linear ramp}[cl][cl]{\color{darkMagenta}Linear ramp with different $\tr$'s}
\psfrag{Bound}[cl][cl]{\color{deepskyblue}Minimum dissipation}
\includegraphics[scale=1]{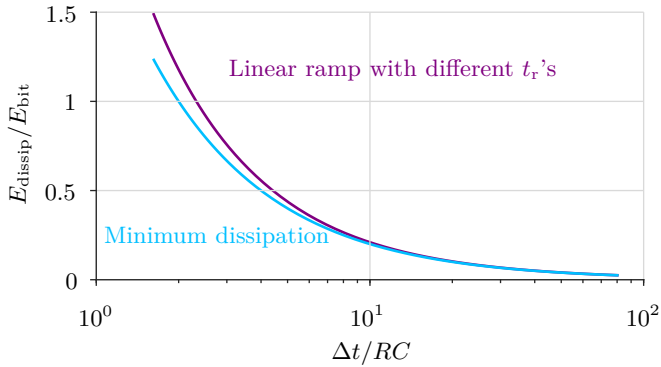}
\caption{Dissipated energy $\Edissip$ versus switching time $\Dt$ for the $RC$ circuit at various speeds. The linear input ramp (of fixed amplitude $\VDD$ but different $\tr$'s.) is compared to the optimal affine protocol.
The switching time $\Dt$ is here defined at $V_1 = \SI{80}{\percent} \, \VDD$.
}
\label{fig_RC_Edissip_vs_Dt_ramp_lin}
\end{figure}

The minimum dissipation \eqref{eq:Edissip RC tr << RC} is naturally obtained with this optimal protocol.
In \cref{fig_RC_Edissip_vs_Dt_ramp_lin}, we compare for different speeds ($\Dt$) the dissipation of the linear ramp (\cref{fig_ramp}) with the lower bound achieved with the affine input (\cref{fig_RC_affine}).
Only the quasi-static regime ($\Dt \gg RC$) was assessed by \cite{Orlov2012,Snider2012}.
Interestingly, we show, by comparison with the minimum-dissipation protocol, that the 
linear ramp deviates from optimality as the speed increases.

\subsection{The nMOS Pass gate}
\label{subsection:CMOS}

The charge transfer process from the supplying input to the capacitive load can be enabled or disabled at will provided that the driving device (abstractly represented by a box is \cref{fig_statement}) is a controlled switch.
This is the primary digital application of the MOS transistor.
We will thus firstly study the energy efficiency of the switching operation through an \emph{nMOS pass gate}, which constitutes a realistic and important nonlinear 
practical 
case of our lower bound \eqref{eq:Edissipmin C} for the dissipation.



\begin{figure}
\centering
\begin{circuitikz}[american voltages, transform shape, line cap=rect, nodes={line cap=butt},scale=0.9]
\draw[black]
(0,0) node[nmos,rotate = -90,yscale=-1] (nMOS) {}
(nMOS.D) to[short,-o] ++(0,0) node[label={[font=\myfontsize,color=black]above:D}] (nD) {}
(nMOS.G) to[short,-o] ++(0,0) node[label={[font=\myfontsize,color=black,yshift=-0.1cm]above:$\VG = \VDD$}] (nG) {}
(nMOS.S) to[short,-o] ++(0,0) node[label={[font=\myfontsize,color=black]above:S}] (nS) {}
;
\draw 
(nMOS.D) -- ++(-\xspacing*\dx,0) node[] (IN) {};
\draw 
(IN) ++ (0,-2*\dx) node[ground,scale=1,color=black] (GNDIN) {}
to[V=\textcolor{b}{$\vIN(t)$},invert] (IN);
\draw 
(nMOS.S) ++(\xspacing*\dx,0) node[label={[font=\myfontsize]above:\textcolor{r}{$\vC(t)$}}] (OUT) {}
(nMOS.S) to[short,-*] (OUT);
\draw
(OUT)++(0,-2*\dx) node[ground,scale=1,color=black] {} to[C,color=black,l_={$C$}] (OUT);
{
\color{mybrown}
\ctikzset{current/distance = 0.5}
\draw 
(nMOS.D)++(0,-0.3*\dx) node[] (nDi) {}
(nMOS.S)++(0,-0.3*\dx) node[] (nSi) {};
\draw[->] 
(nDi) -- node[below, color=mybrown, font=\myfontsize] {$i(t)$} (nSi);
}
\end{circuitikz}
\caption{nMOS pass gate driving a capacitor.}
\label{fig_pass_gate}
\vspace{2mm}
\centering
\psfragscanon
\footnotesize
\psfrag{t [ns]}[cc][cc]{\normalsize $t \, [\si{\nano\second}]$}
\psfrag{v(t) [V]}[cc][cc]{\normalsize  $v(t) \, [\si{\volt}]$}
\psfrag{i(t)/dv(t) [uS]}[cc][cc]{\normalsize $\textcolor{orange}{i(t)/\vDS(t)} \, [\si{\micro\siemens}]$}
\psfrag{0}[cc][cc]{$0$}
\psfrag{0.5}[cc][cc]{$0.5$}
\psfrag{0.8}[cc][cc]{$0.8$}
\psfrag{1}[cc][cc]{$1$}
\psfrag{10}[cc][cc]{$10$}
\psfrag{20}[cc][cc]{$20$}
\psfrag{30}[cc][cc]{$30$}
\psfrag{40}[cc][cc]{$\Dt$}
\psfrag{50}[cc][cc]{$50$}
\psfrag{100}[cc][cc]{$100$}
\psfrag{200}[cc][cc]{$200$}
\psfrag{vIN(t)}[tl][tl]{\small $\textcolor{b}{\vIN(t)}$}
\psfrag{vOUT(t)}[tl][tl]{\small $\textcolor{r}{\vC(t)}$}
\psfrag{vDS(t)}[tl][tl]{\small $\textcolor{m}{\vDS(t)}$}
\psfrag{p(t)}[tl][tl]{\small $\textcolor{darkMagenta}{p_{\mathrm{dissip}}(t)} = \textcolor{m}{\vDS(t)} \textcolor{mybrown}{i(t)}$}
\psfrag{a.VDD}[bl][bl]{\small $V_1 = \alpha \, \VDD$}
\psfrag{tr}[tc][tc]{\small  $\textcolor{b}{\tr}$}
\psfrag{I}[bc][bc]{I}
\psfrag{II}[bc][bc]{II}
\includegraphics[scale=1]{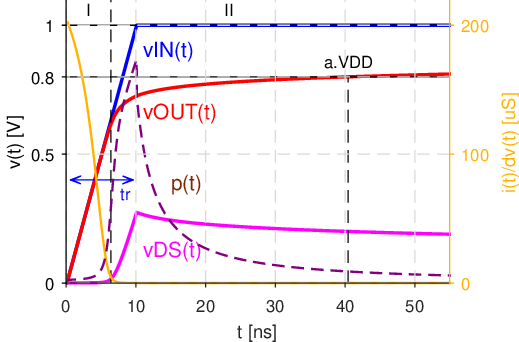}
\caption{Simulated pass gate: $\SI{28}{\nano\meter}$ FD-SOI regular-$\Vth$ nMOS of minimal dimensions $L = \SI{30}{\nano\meter}$ and $W = \SI{80}{\nano\meter}$;
$\VDD = \SI{1}{\volt}$;
$C = \SI{1}{\femto\farad}$;
input rise time $\tr = \SI{10}{\nano\second}$.
\newline
Logical 1 definition (defining the bit-switching time): \mbox{$V_1 = \SI{80}{\percent} \, \VDD$}.
\newline
The dissipated power $p_{\mathrm{dissip}}(t)$ (in magenta) is plotted in arbitrary units.
}
\label{fig_pass_gate_tr_10ns}
\end{figure}

A pass gate, implemented with an nMOS transistor, is connected to a constant capacitor in \cref{fig_pass_gate}.
The charging process is ensured by a linear input voltage ramp (as previously sketched in \cref{fig_ramp}) whose rise time $\tr$ can vary by several orders of magnitude.
$\VDD$ is the supply voltage of the used CMOS technology.

For this realistic case, we avoid to resort to a simplified analytical model of the transistor that would be highly inaccurate and unfaithful to the small-dimension effects in the most advanced technologies \cite{Tsividis2011} favoured for digital circuit design \cite{Baker}. 
Instead, in the same spirit as \cite[Section V]{Snider2012}, we have performed SPICE simulation
in the time domain, compatible with the industrial process design kits. 
We emphasize that methodology and our main result \eqref{eq:Edissipmin} do not require any analytical model of the studied circuit, even though it was first illustrated for the linear $RC$ circuit.

\subsubsection{Transient Simulation}


The simulated waveforms in switching conditions are depicted in \cref{fig_pass_gate_tr_10ns}, using a colour code consistent with \cref{fig_pass_gate}. 
The voltage defining the logic level 1 is $V_1 = \SI{0.8}{\volt}$ and the response time $\Dt$ is extracted as the time needed for $\vC(t)$ to reach at least $V_1$, about $\SI{40}{\nano\second}$ for the illustrated case.
We further observe that the nMOS is incapable to fully pass the logical 1 to be written on the capacitor. Although this behaviour is well known \cite{Baker}, we will revisit and reinterpret it with the introduced notions of instantaneous and average conductances \eqref{eq:Gav} that appears in the minimum-dissipation formula \eqref{eq:Edissipmin C}.

The nMOS transistor is a strongly nonlinear device, in the sense that the current flowing from the drain (by convention input of the pass gate in \cref{fig_pass_gate}) to the source (output) is a nonlinear function of its terminal voltage differences ($\vGS$, and $\vDS \equiv \Dv$), from strong to weak inversion \cite{Tsividis2011}. Consequently, the conductance of the transistor (defined as $i(t)/\Dv(t)$) varies a lot during the bit writing operation.
We conceptually distinguish two phases in \cref{fig_pass_gate_tr_10ns}.

In the first (I) phase,  the charging process is efficient.
The $\vGS(t) = \VDD - \textcolor{k}{\vC(t)}$ is large ($\gg \Vth$, the threshold voltage of the transistor), meaning that the MOS is in \emph{strong inversion} (large conductance $i(t)$).
Thanks to the driving capability of the device, $\textcolor{k}{\vC(t)}$ closely follows  $\textcolor{k}{\vIN(t)}$ (similarly to \cite{Orlov2012}'s experiment on the RC circuit in the quasi-static switching conditions discussed in \cref{subsection:RC}).
As long as the 
$\Dv(t)$
 remains small (the transistor is in linear region), the instantaneous dissipated power $p_{\mathrm{dissip}}(t)$ is low.

Then, in the second (II) phase, the charge transfer becomes more and more inefficient. 
The $\vGS(t)$ gets close or even below the $\Vth$, i.e. the transistor operates in weak inversion (low conductance), and $\textcolor{k}{\vC(t)}$ painfully follows $\textcolor{k}{\vIN(t)}$ (“the nMOS is not good at passing a 1” \cite{Baker}).
The larger $\Dv(t)$ (see \cref{fig_pass_gate_tr_10ns}) increases the dissipation, which experiences a peak at $\SI{10}{\nano\second}$.
Although $\vC(t)$ already reaches more than $\SI{60}{\percent} \, \VDD$ in the first phase (while the final objective is $\SI{80}{\percent} \, \VDD$), the second phase
of the charging process dominates the energy dissipation of the bit writing operation.

\subsubsection{Analysis of the Dissipation}

\begin{figure}
\centering
\psfragscanon
\psfrag{e0}[cc][cc]{\footnotesize $10^0$}
\psfrag{em1}[cc][cc]{\footnotesize $10^{-1}$}
\psfrag{em2}[cc][cc]{\footnotesize $10^{-2}$}
\psfrag{em3}[cc][cc]{\footnotesize $10^{-3}$}
\psfrag{em4}[cc][cc]{\footnotesize $10^{-4}$}
\psfrag{em5}[cc][cc]{\footnotesize $10^{-5}$}
\psfrag{em6}[cc][cc]{\footnotesize $10^{-6}$}
\psfrag{em7}[cc][cc]{\footnotesize $10^{-7}$}
\psfrag{em8}[cc][cc]{\footnotesize $10^{-8}$}
\psfrag{DT [s]}[cc][cc]{$\Dt\, [\si{\second}]$}
\psfrag{Edissip/(C.V1^2/2)}[bc][cc]{$\Edissip/\EC$}
\psfrag{Linear ramp}[bl][bl]{\color{darkMagenta}Actual dissipation (linear ramp)}
\psfrag{Minimum dissipation}[bl][bl]{\color{deepskyblue}Minimum dissipation}
\includegraphics[scale=1]{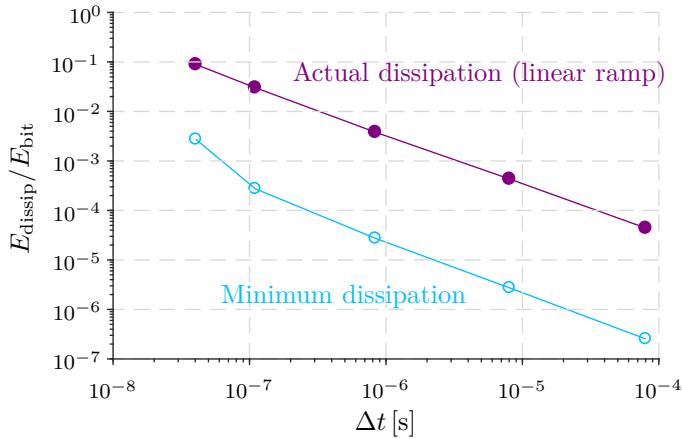}
\caption{Capacitance bit writing through an nMOS pass gate (same parameters as \cref{fig_pass_gate_tr_10ns}, but different $\tr$'s): dissipation versus response time $\Dt$.
All the quantities were extracted from SPICE circuit simulations.
}
\label{fig_pass_gate_Edissip}
\end{figure}

In \cref{fig_pass_gate_Edissip}, we have plotted the dissipation, extracted from the transient simulations performed for different $\tr$, as a function of the response time (that must be also be empirically extracted for each $\tr$). We again observe, this time for the nonlinear MOS device, the trend $\Edissip \propto 1/\Dt$, widely reported for the linear $RC$ circuit (see \cref{subsection:RC} and associated references).
The fact that the total dissipation is quite high ($\SI{10}{\percent}$ of the bit energy $C V_1^2/2$ for the fastest case of \cref{fig_pass_gate_tr_10ns}) is attributed to the poor conductance of the transistor in the second phase of the charging process discussed here above.

To the actual dissipation we compare the minimum dissipation of the form \cref{eq:Edissipmin C} (constant capacitance load but general nonlinear driving device).
For each simulated case, the average conductance is calculated according to \eqref{eq:Gav} and all the quantities of \cref{eq:Edissipmin C} are then known.
We remind that \eqref{eq:Edissipmin C} predicts the lowest dissipation that could, in theory, be achieved with the optimal protocol $\vIN(t)$ guaranteeing the criterion \eqref{eq:Dv(t) cst} (not determined in this first paper and left for further work).
The discrepancy with the exact dissipation is more than two orders of magnitude which reveals that as of it, the linear profile for $\vIN(t)$ results in a highly inefficient switching process in terms of dissipation.

\subsection{The CMOS Transmission Gate}

\begin{figure}
\centering
\begin{circuitikz}[american voltages, transform shape, line cap=rect, nodes={line cap=butt},scale=0.9]
\draw[black]
(0,0) node[nmos,rotate = -90,yscale=-1] (nMOS) {}
(nMOS.D) to[short,-o] ++(0,0) node[label={[font=\myfontsize,color=black]left:IN}] (nD) {}
(nMOS.G) to[short,-o] ++(0,0) node[label={[font=\myfontsize,color=black,yshift=-0.1cm]above:$\VDD$}] (nG) {}
(nMOS.S) to[short,-o] ++(0,0) node[label={[font=\myfontsize,color=black]right:OUT}] (nS) {}
(0,0) node[pmos,rotate = +90,yscale=-1] (pMOS) {}
(pMOS.G) to[short,-o] ++(0,0) node[label={[font=\myfontsize,color=black,yshift=+0.1cm]below:$0$}] (nGp) {}
;
\end{circuitikz}
\caption{CMOS transmission gate.}
\label{fig_transmission_gate}
\end{figure}
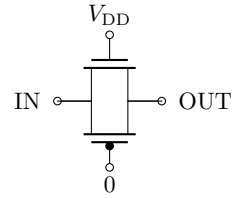
\begin{figure}
\centering
\psfragscanon
\psfrag{e0}[cc][cc]{\footnotesize $10^0$}
\psfrag{em1}[cc][cc]{\footnotesize $10^{-1}$}
\psfrag{em2}[cc][cc]{\footnotesize $10^{-2}$}
\psfrag{em3}[cc][cc]{\footnotesize $10^{-3}$}
\psfrag{em4}[cc][cc]{\footnotesize $10^{-4}$}
\psfrag{em5}[cc][cc]{\footnotesize $10^{-5}$}
\psfrag{em6}[cc][cc]{\footnotesize $10^{-6}$}
\psfrag{em7}[cc][cc]{\footnotesize $10^{-7}$}
\psfrag{em8}[cc][cc]{\footnotesize $10^{-8}$}
\psfrag{em9}[cc][cc]{\footnotesize $10^{-9}$}
\psfrag{em10}[cc][cc]{\footnotesize $10^{-10}$}
\psfrag{em11}[cc][cc]{\footnotesize $10^{-11}$}
\psfrag{DT [s]}[cc][cc]{$\Dt\, [\si{\second}]$}
\psfrag{Edissip/(C.V1^2/2)}[bc][cc]{$\Edissip/\EC$}
\psfrag{Linear ramp}[bl][bl]{\color{darkMagenta}Actual dissipation}
\psfrag{Minimum dissipation}[bl][bl]{\color{deepskyblue}Minimum dissipation}
\psfrag{C*VDD^2/2}[bl][bl]{\color{r}\small$C \, \VDD^2 /2$}
\psfrag{C*V1^2/2}[tl][tl]{\color{k}\small$\EC$}
\includegraphics[scale=1]{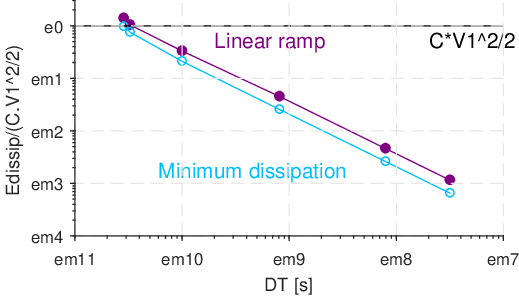}
\caption{bit writing through a CMOS transmission gate with a linear ramp (same transistor dimensions and parameters as \cref{fig_pass_gate_tr_10ns}; $\tr$ range from $\SI{1}{\pico\second}$ to $\SI{40}{\nano\second}$): dissipation versus response time $\Dt$.
}
\label{fig_transmission_gate_Edissip}
\end{figure}

The CMOS transmission gate was introduce to reduce the shortcoming of the pass gate highlighted above, as our bound allows to quantify.

The schematic of the CMOS transmission gate is provided in \cref{fig_transmission_gate}. The complementary pMOS transistor added in parallel with the nMOS overcomes the difficulty of fully passing a logical 1. When $\vC(t)$ becomes larger than $\VDD/2$ and closer to $\VDD$, the conductance of the pMOS increases and it takes over from the nMOS for further charge transfer.
Since there is always one of the two transistors which is highly conductive instantaneously, the charging process remains efficient all the way. The bit-switching time is reduced, compared to the single pass gate of \cref{fig_pass_gate}, and so it the energy dissipation.

The dissipation is depicted in \cref{fig_transmission_gate_Edissip} for different speeds.
For a response time of $\Dt = \SI{40}{\nano\second}$ (same as discussed for the nMOS pass gate in \cref{fig_pass_gate_tr_10ns,fig_pass_gate_Edissip}), we can find $\Edissip \approx \num{e-3} \,\EC$ in \cref{fig_transmission_gate_Edissip}, i.e. a reduction of the dissipation by a factor \num{100}.
When the speed is further increases by using smaller $\tr$ for the linear ramp, the dissipation increases accordingly.
There is a minimum write time $\Dt \approx \SI{30}{\pico\second}$ for given supply voltage $\VDD$ and load $C$.

The energy efficiency of the CMOS transmission gate is highlighted by our bound \eqref{eq:Edissipmin C}, also extracted and represented in \cref{fig_transmission_gate_Edissip}. For all the speed, the ratio $\Edissip/\Edissipmin$ \eqref{eq:Edissip/EC C} is smaller than $2$.
Such result is a consequence of the excellent switching capability of the transmission gate, even when the input signal is a simple linear ramp.
This gap could be narrowed even further by resorting to an optimal protocol.

Finally, it is interesting to observe that the actual dissipation \emph{exceeds} $\EC$ for the fastest cases, that is the yield of the energy transfer defined in \cref{eq:EC/EIN} falls below $\SI{50}{\percent}$. This means that of the energy supplied by the input, less than $\SI{50}{\percent}$ is stored as a reversible energy in the capacitor whereas more than $\SI{50}{\percent}$ are wasted as heat by the transmission gate.

\section{Discussion and Conclusions}
\label{Discussion and Conclusions}

Revisiting \cite{Orlov2012}'s experimental results, we highlighted the fact that the actual dissipation can be computed analytically and tightly matches both the experimental results and the prediction of the novel,  “non-Landauer”  fundamental bound that we have derived and introduced.
We have also addressed the non-quasi-static (fast input) case. Moreover, we investigated realistic CMOS pass and transmission gates, through 
SPICE simulations that are compared to our theoretical bound.
The observed discrepancies are consistent with our knowledge of the MOS transistor behaviour and their limitations. 
Our fundamental bound can be regarded as a tool to compare the energy efficiency different bit-switching architectures and protocols, even across various technologies and not only CMOS thanks its generality.
We hope to provide further application examples in future work, such as diode, quantum dots \cite{Lent2006} or adiabatic logic gates \cite{Athas1994,Ye1996,Snider2012,Frank2020}.

Moreover, our “non-Landauer” lower bound, in its most general form \eqref{eq:Edissipmin}, not dependent on the linear capacitive load: it predicts the energy cost of passing a certain amount of charge in a certain amount of time through an arbitrary dissipative device.
Although we have illustrated it on bit writing operation, where the bit is recorded as a voltage level in a linear capacitance, we wish to generalise to arbitrary devices, with multiple inputs and outputs as well as with a dynamic internal behaviour. 
We believe the port-Hamiltonian framework \cite{Gernandt2021} is suitable toward that goal. 
Applications could 
include more complex circuits such as SRAM bitcells (dynamic nonlinear capacitive load).

%


\bibliographystyle{IEEEtran}

\bibliography{IEEEabrv,bib}

\begin{thebibliography}{10}
\providecommand{\url}[1]{#1}
\csname url@samestyle\endcsname
\providecommand{\newblock}{\relax}
\providecommand{\bibinfo}[2]{#2}
\providecommand{\BIBentrySTDinterwordspacing}{\spaceskip=0pt\relax}
\providecommand{\BIBentryALTinterwordstretchfactor}{4}
\providecommand{\BIBentryALTinterwordspacing}{\spaceskip=\fontdimen2\font plus
\BIBentryALTinterwordstretchfactor\fontdimen3\font minus
  \fontdimen4\font\relax}
\providecommand{\BIBforeignlanguage}[2]{{%
\expandafter\ifx\csname l@#1\endcsname\relax
\typeout{** WARNING: IEEEtran.bst: No hyphenation pattern has been}%
\typeout{** loaded for the language `#1'. Using the pattern for}%
\typeout{** the default language instead.}%
\else
\language=\csname l@#1\endcsname
\fi
#2}}
\providecommand{\BIBdecl}{\relax}
\BIBdecl

\bibitem{Landauer1961}
R.~Landauer, ``{Irreversibility and Heat Generation in the Computing
  Process},'' \emph{IBM journal of research and development}, vol.~5, no.~3,
  pp. 183--191, 1961.

\bibitem{Izydorczyk2009}
J.~Izydorczyk, ``{Three Steps to the Thermal Noise Death of Moore’s Law},''
  \emph{IEEE transactions on very large scale integration (VLSI) systems},
  vol.~18, no.~1, pp. 161--165, 2009.

\bibitem{Parrondo2015}
J.~M. Parrondo, J.~M. Horowitz, and T.~Sagawa, ``Thermodynamics of
  information,'' \emph{Nature physics}, vol.~11, no.~2, pp. 131--139, 2015.

\bibitem{Orlov2012}
A.~O. Orlov, C.~S. Lent, C.~C. Thorpe, G.~P. Boechler, and G.~L. Snider,
  ``{Experimental Test of Landauer's Principle at the Sub-{$k_{\mathrm{B}} T$}
  Level},'' \emph{Japanese Journal of Applied Physics}, vol.~51, no.~6S, p.
  06FE10, 2012.

\bibitem{Snider2012}
G.~L. Snider, E.~P. Blair, C.~C. Thorpe, B.~T. Appleton, G.~P. Boechler, A.~O.
  Orlov, and C.~S. Lent, ``{There is No Landauer Limit: Experimental Tests of
  the Landauer Principle},'' in \emph{2012 12th IEEE International Conference
  on Nanotechnology (IEEE-NANO)}.\hskip 1em plus 0.5em minus 0.4em\relax IEEE,
  2012, pp. 1--6.

\bibitem{Athas1994}
W.~C. Athas, L.~J. Svensson, J.~G. Koller, N.~Tzartzanis, and E.~Y.-C. Chou,
  ``{Low-Power Digital Systems Based on Adiabatic-Switching Principles},''
  \emph{IEEE Transactions on Very Large Scale Integration (VLSI) Systems},
  vol.~2, no.~4, pp. 398--407, 1994.

\bibitem{Ye1996}
Y.~Ye and K.~Roy, ``{Energy Recovery Circuits Using Reversible and Partially
  Reversible Logic},'' \emph{IEEE Transactions on Circuits and Systems I:
  Fundamental Theory and Applications}, vol.~43, no.~9, pp. 769--778, 1996.

\bibitem{Frank2020}
M.~P. Frank, R.~W. Brocato, B.~D. Tierney, N.~A. Missert, and A.~H. Hsia,
  ``{Reversible Computing with Fast, Fully Static, Fully Adiabatic CMOS },'' in
  \emph{2020 International Conference on Rebooting Computing (ICRC)}.\hskip 1em
  plus 0.5em minus 0.4em\relax IEEE, 2020, pp. 1--8.

\bibitem{Tsividis2011}
Y.~Tsividis and C.~McAndrew, \emph{{Operation and Modeling of the MOS
  Transistor}}, ser. The Oxford Series in Electrical and Computer Engineering
  Series.\hskip 1em plus 0.5em minus 0.4em\relax Oxford University Press, 2011.

\bibitem{Baker}
R.~J. Baker, \emph{{CMOS Circuit Design, Layout, and Simulation}},
  3rd~ed.\hskip 1em plus 0.5em minus 0.4em\relax John Wiley \& Sons, IEEE
  Press, 2010.

\bibitem{Lent2006}
C.~S. Lent, M.~Liu, and Y.~Lu, ``Bennett clocking of quantum-dot cellular
  automata and the limits to binary logic scaling,'' \emph{Nanotechnology},
  vol.~17, no.~16, p. 4240, 2006.

\bibitem{Gernandt2021}
H.~Gernandt, F.~Haller, T.~Reis, and A.~J. van~der Schaft, ``{Port-Hamiltonian
  formulation of nonlinear electrical circuits},'' \emph{Journal of Geometry
  and Physics}, vol. 159, p. 103959, 2021.

\end{thebibliography}

\appendices

\crefalias{section}{appendix}
\crefalias{subsection}{appendix}

\section{Lower Bound for the Dissipation}
\label{appendix:Edissipmin}


We prove our main result \eqref{eq:Edissipmin} with the  \emph{Cauchy–Schwarz inequality}, applied to a special scalar product that is the integral of the product of two functions:
\begin{equation}
\label{eq:Cauchy}
\bigg( \int f(t) \, g(t) \, \dt \Big)^2 \leq \int \big(f(t)\big)^2 \, \dt \cdot \int \big(g(t)\big)^2 \, \dt
\text{.}
\end{equation}
Let $f(t) \equiv \sqrt{i(t)/\Dv(t)}$ and $g(t) \equiv \sqrt{i(t)\Dv(t)}$. The direct application of \eqref{eq:Cauchy} provides
\begin{equation}
\label{eq:Cauchy proof}
\begin{aligned}
\bigg( \int_{t_0}^{t_0 \mathrlap{+ \Dt}} i(t) \, \dt \bigg)^2 
& = \bigg( \int_{t_0}^{t_0 + \Dt} \! \sqrt{\frac{i(t)}{\Dv(t)}}
\cdot \sqrt{i(t) \Dv(t)} 
 \, \dt \bigg)^2 \\
& \leq 
\underbrace{\int_{t_0}^{t_0 + \Dt} \! \frac{i(t)}{\Dv(t)} \, \dt
}_{\ds \equiv \Dt \, \Gav}
\ \cdot \  
\underbrace{\int_{t_0}^{t_0 \mathrlap{+ \Dt}} i(t) \Dv(t) \, \dt
}_{\ds \vphantom{\equiv \Dt \, \Rav}= E_{\mathrm{dissip}}}
\end{aligned}
\end{equation}
We identify the $\Dq$ 
defined in \eqref{eq:Dq} and we rewrite \eqref{eq:Cauchy proof} as \eqref{eq:Edissipmin}, the main result of this paper.

The Cauchy-Schwarz inequality (thus our lower bound) is \emph{tight} is lower bound if and only if the vectors $f(t)$ and $g(t)$ are collinear, i.e. $\sqrt{i(t)/\Dv(t)} \propto \sqrt{i(t)\Dv(t)}$ or
$
\Dv(t) = \Delta V = \text{constant}
\text{.}
$


\end{document}